\documentclass[letter, amsfonts, amssymb, amsmath, reprint, showkeys, 
twoside, twocolumn,
elsarticle,
superscriptaddress]{revtex4-1}

\usepackage{xcolor}
\usepackage{graphicx} 
\usepackage{subfigure}
\usepackage{dcolumn} 
\usepackage{bm}
\usepackage{float}
\usepackage[columnwise,running]{lineno}
\usepackage[normalem]{ulem}
\usepackage{slashed}

\usepackage{hyperref}
\usepackage{setspace}
\usepackage{color}
\definecolor{orange}{cmyk}{0.,0.353,1.,0.}    % orange

\usepackage{tikz}
\usetikzlibrary{positioning}
\begin{document}

\title{Charge Separation Measurements in Au+Au collisions at $\sqrt{s_{NN}}=$ 7.7--200 GeV
in Search of the Chiral Magnetic Effect 
 }

\author{The STAR Collaboration}

\begin{abstract}
%\onehalfspacing
The chiral magnetic effect in heavy-ion collisions predicts a 
charge separation signal along a magnetic field, which indicates local $\cal{P}$ and $\cal{CP}$ violations in the quark-gluon plasma.
We report measurements of electric charge separation signals perpendicular to the spectator event plane in Au+Au collisions using high-statistics data from RHIC Beam Energy Scan II and top-energy runs.
A novel event shape selection method is employed to suppress the flow-induced background. The residual charge separation signals near the zero-flow limit are positive in Au+Au collisions within the 20\%--50\% centrality range, with significance levels of $2.6\sigma$, $3.1\sigma$, and $3.3\sigma$ at $\sqrt{s_{NN}} =$ 11.5, 14.6, and 19.6 GeV, respectively. At other beam energies, the signals are either statistically limited or consistent with zero.

\begin{description}
\item[keywords]
heavy-ion collision; chiral magnetic effect; event shape selection
\end{description}
\end{abstract}
\maketitle

The chiral magnetic effect (CME) describes the induction of an electric current ($\vec{J}$) along the direction of an intense magnetic field ($\vec{B}$) due to a finite chirality chemical potential of fermions ($\mu_5 \neq 0$)
in a quark-gluon plasma (QGP)~\cite{CME-1,CME-2,CME-3} or a condensed-matter system. 
This phenomenon, $\vec{J} \propto \mu_5\vec{B}$, has been experimentally confirmed in systems such as Dirac~\cite{cme-nature,cme-science} and Weyl~\cite{cme-weyl} semimetals, where $\mu_5$ and $\vec{B}$ can be precisely controlled.
The CME in high-energy heavy-ion collisions is more elusive~\cite{kharzeev-liao, Baryogenesis, STAR-BESI-mini-review, Kharzeev:2024zzm}, as its formation requires synchronizing several preconditions in a fast-evolving system and its detection demands eliminating intricate backgrounds. 
Nonetheless, searching for the CME in quarks could provide insights into the fundamental aspects of quantum chromodynamics (QCD), such as topological vacua~\cite{CME-1,CME-2,CME-3}, which have yet to be experimentally observed.

Heavy-ion collisions at the BNL Relativistic Heavy Ion Collider (RHIC) and the CERN Large Hadron Collider (LHC) can create a deconfined QGP, whose physical conditions are theorized to have existed microseconds after the Big Bang~\cite{bigbang}. Chiral symmetry restoration is anticipated to accompany deconfinement, whereby light quarks become nearly massless and bear definite chirality.
A possible chirogenesis mechanism suggests that collective gluon excitations generate topological vacuum transitions, transferring net chirality to (anti)quarks within the rapidly expanding QGP droplets~\cite{CME-1, CME-2, CME-3}.
The $\mathcal{P}$- and $\mathcal{CP}$-violating chirogenesis in the strong interaction is a prerequisite for the CME, and if observed, could serve as an analogous mechanism for understanding baryogenesis in the electroweak interaction of the early Universe~\cite{Baryogenesis}.

\begin{figure}[hpbt]
\centering
\includegraphics[width=0.4\textwidth]{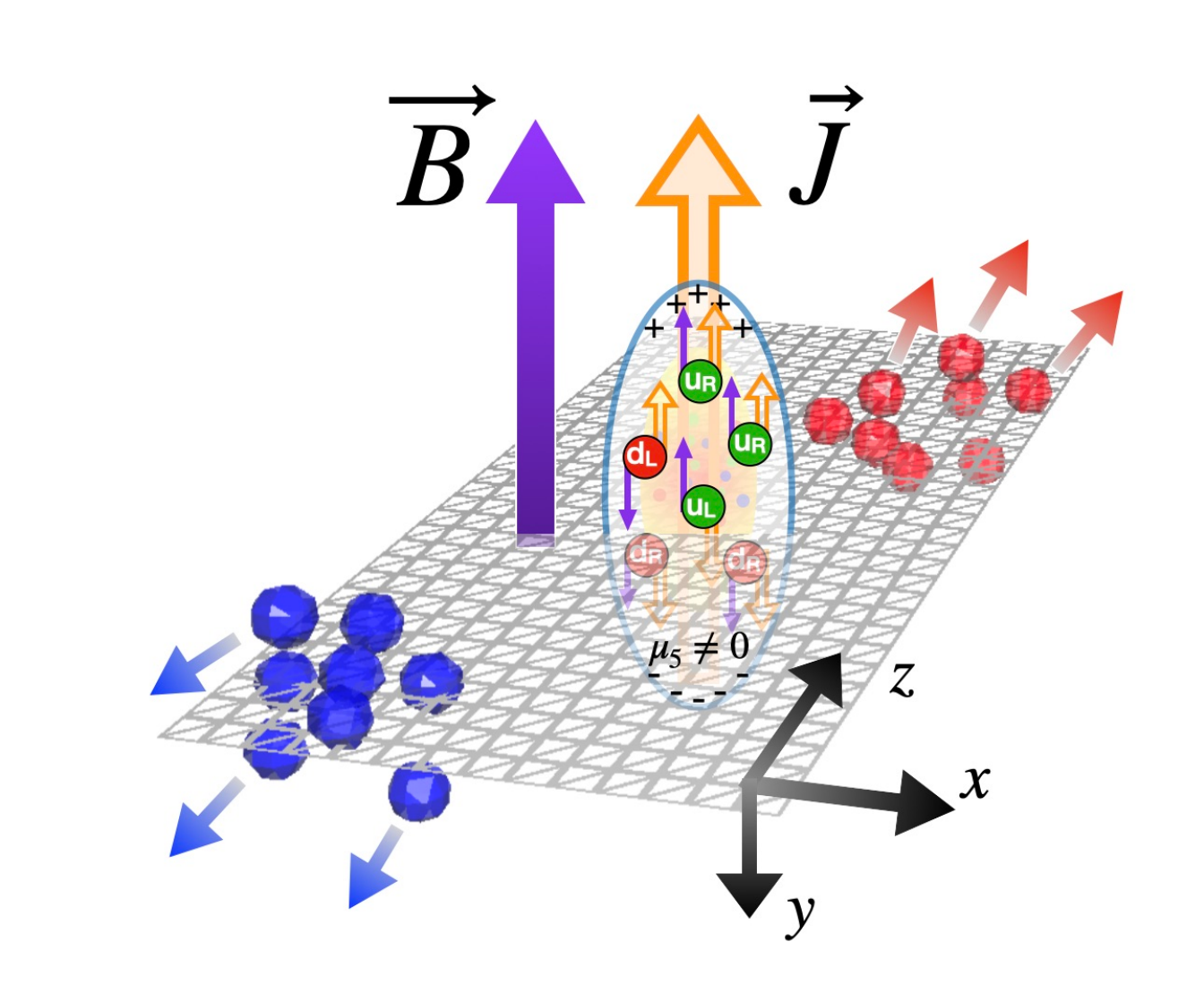}
\caption{Sketch of the CME in a two-nucleus collision. The forward- and backward-moving nucleon groups denote spectator nucleons that do not participate in the collision but contribute to a strong magnetic field $\vec{B}$. In the participant zone (the overlap region), the  CME induces an electric current ($\vec{J}$) across the $x$-$z$ plane in chiral domains ($\mu_5\neq 0$). 
}
\label{fig1}
\end{figure}

Figure~\ref{fig1} illustrates spectator nucleons passing each other at nearly the speed of light in noncentral collisions. 
The resulting magnetic field, perpendicular to the reaction plane (RP, the $x$-$z$ plane), is estimated to peak around $10^{17}-10^{19}$ 
Gauss~\cite{CME-5,B-field-1,B-field-2}.
With nonzero $\mu_5$ in the participant zone, the CME could thus yield an electric charge separation orthogonal to the RP from (anti)quarks in metastable domains. 
Though the transient magnetic field is among the strongest in the cosmos, its impact on final-state particles is difficult to detect due to its rapid decay. 

The charge separation may be quantified with Fourier coefficients of the azimuthal angle ($\varphi$) distribution of final-state charged particles ($N_\pm$) with respect to the RP ($\Psi_{\rm RP}$)~\cite{CME-8},
\begin{equation}
\frac{dN_{\pm}}{d\varphi} \propto 1 + \sum_{n=1}^\infty 2v_{n}^{\pm}\cos n\Delta\varphi + 2a_{1}^{\pm}\sin\Delta\varphi,
\end{equation}
where $\Delta\varphi = \varphi-\Psi_{\rm{RP}}$. 
Hydrodynamic expansion~\cite{hydro} of the system converts the initial overlap geometry into a momentum anisotropy ($v_n$) of the final-state particles.
The $a_{1}^{+}$ and $a_{1}^{-}$ bear opposite signs because of charge separation and are on the order of one over multiplicity~\cite{CME-8}.
Since parity can be locally violated but globally conserved in QCD, 
a CME observable aims to measure the fluctuation of $a_1$.  
A recent study~\cite{CME-9} affirms the similarity in the key components of different CME-sensitive observables~\cite{CME-8, Magdy, Tang}.  
We focus on the most widely used $\gamma^{112} \equiv \langle \cos (\varphi_\alpha + \varphi_\beta - 2\Psi_{\rm{RP}}) \rangle$~\cite{CME-8}, where $\alpha$ and $\beta$ represent charge sign, and report $\Delta\gamma^{112} \equiv \gamma_{\rm{OS}}^{112} - \gamma_{\rm{SS}}^{112}$, which cancels the common background for opposite-sign (OS) and same-sign (SS) pairs.

A CME signal ($\approx 2|a_1^\pm|^2$) contributes positively to $\Delta\gamma^{112}$, while elliptic flow ($v_2$) forms a major background by coupling with  other mechanisms,
including resonance decays~\cite{CME-8}, local charge conservation (LCC)~\cite{PrattSorren:2011,LCC-1}, and transverse momentum conservation (TMC)~\cite{Pratt2010,Flow_CME}.
The flow-related background complicates the interpretation of the positive $\Delta\gamma^{112}$ data observed in Au+Au collisions at the top RHIC energy~\cite{STAR-1,STAR-2,STAR-3} and in Pb+Pb collisions at the LHC energies~\cite{ALICE-1}.
Additionally, nonflow correlations~\cite{STAR-v2-first,STAR-5} from various sources, including clusters, resonances, jets, and dijets, are unrelated to the RP but can also contribute positively to $\Delta\gamma^{112}$.
Nonflow effects can be minimized by reconstructing the RP with spectator information~\cite{STAR:2005btp}.
One recent attempt to mitigate the flow-related background involves discerning subtle signal differences between two isobaric systems with similar backgrounds, $^{96}_{44}$Ru+$^{96}_{44}$Ru and $^{96}_{40}$Zr+$^{96}_{40}$Zr~\cite{isobar_proposal1,isobar_proposal2,isobar_proposal3}.
A CME signal was not observed in STAR isobar data at $\sqrt{s_{NN}} = 200$ GeV ~\cite{Isobar-1,Isobar-2}, indicating that the background contributions dominate over any potential signal at this center-of-mass energy.

As the signal fraction of $\Delta\gamma^{112}$ may be too small in small collision systems at $\sqrt{s_{NN}} = 200$ GeV \cite{ESS-1}, our focus shifts towards larger collision systems with more protons and a range of energies in the RHIC Beam Energy Scan (BES) program during 2010--2014 (I) and 2018--2021 (II).
STAR analyzed the BES-I data using an observable $H(\kappa_{bg})\equiv (\kappa_{bg} v_2 \delta -\gamma^{112})/(1+\kappa_{bg} v_2)$~\cite{STAR-BES-1}, where $\kappa_{bg}$ is an adjustable parameter unknown a priori, and $\delta \equiv \langle \cos(\varphi_\alpha-\varphi_\beta)\rangle$ represents a two-particle correlation.
The $\kappa_{bg} v_2 \delta$ represents the flow-related background.
With $\kappa_{bg} \approx$ 2--3, the difference $\Delta H \equiv H_{\rm SS}-H_{\rm OS}$ in the 0--60\% centrality region of Au+Au collisions tends to vanish at $\sqrt{s_{NN}} = 7.7$ and 200 GeV and remains positive finite at beam energies in between~\cite{STAR-BES-1}.
Recent STAR data of directed flow ($v_1$) unveiled the magnetic field's imprint on the QGP in Au+Au collisions, with the effect much stronger at $\sqrt{s_{NN}} = 27$ GeV than at 200 GeV~\cite{STAR-Bfield}.
This suggests that the opportunity
to detect the CME may be enhanced using the BES-II data.
The BES-II program provides over ten times the statistics of BES-I, with a dedicated detector for reconstructing the spectator plane. 
At sufficiently low energies, the lack of deconfinement also enables testing for the absence of the CME signal.
 
The manifested background is primarily due to the emission pattern of final-state particles, influenced by fluctuations in both the initial overlap geometry (eccentricity) and the later-stage expansion of the QGP~\cite{berndt-2}.
Some methods of event shape engineering~\cite{SergeiESE,ALICE-2,ALICE-3,CMS-2} construct an event shape variable from a kinematic region excluding particles of interest (POI), relying on long-range flow correlations.
However, flow-plane decorrelations could break this long-range flow assumption~\cite{rn-CMS, decor-ATLAS, decor-ATLAS2, Maowu}.
These methods also involve considerable extrapolation towards zero $v_2$, leading to substantial uncertainties~\cite{ESS-2}.
In this work, we adopt a novel event shape selection (ESS) method~\cite{1st-ESS,ESS-1,ESS-2} that captures both geometric and expansion properties while enabling a more precise extrapolation toward zero $v_2$.
The details are discussed in an accompanying paper~\cite{accompany-long-paper}.

\begin{figure}[bt]
\centering
\includegraphics[width=0.48\textwidth]{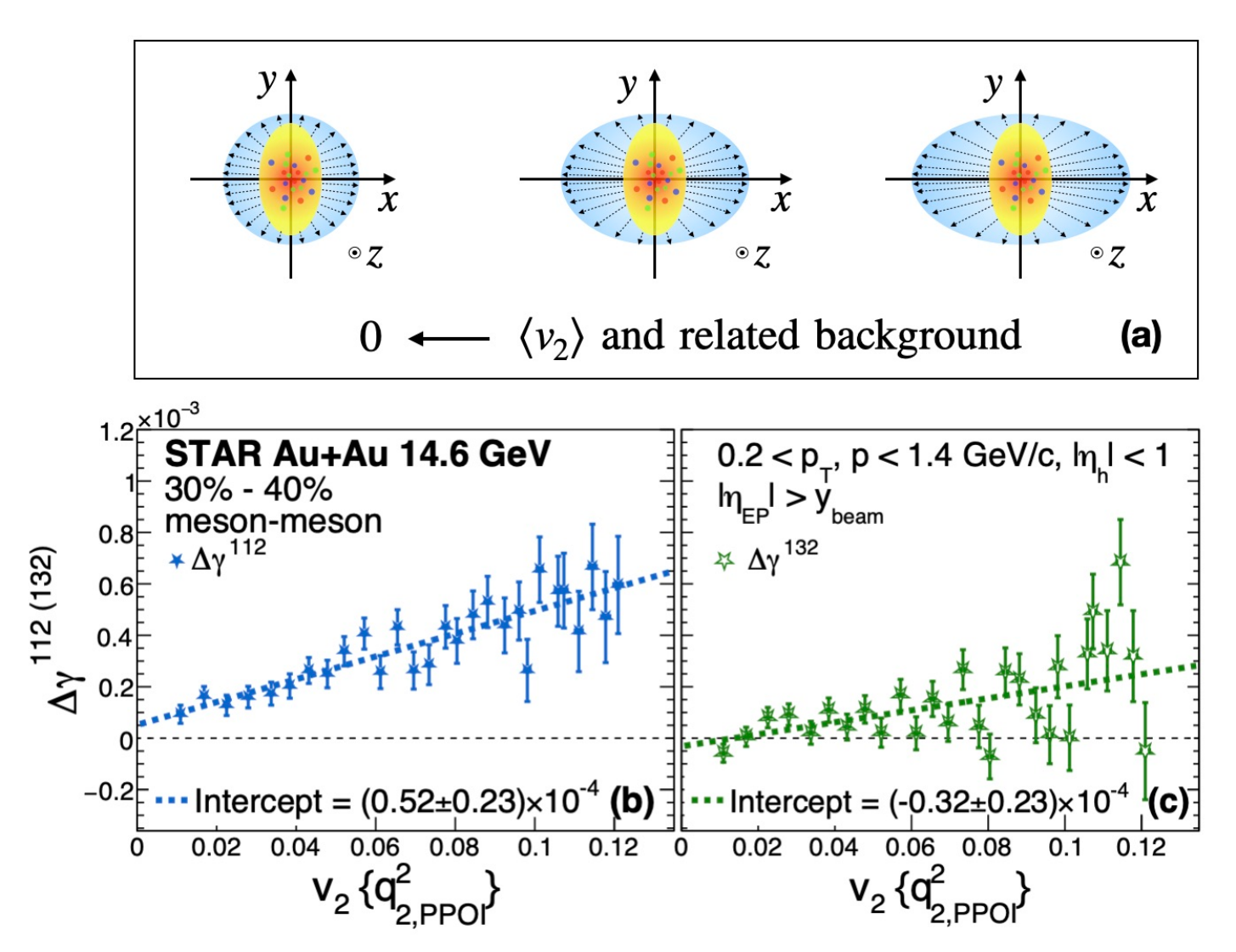}
\caption{(a): illustration of various particle emission patterns (outer regions spanned by arrows) for different events with the same initial geometry (inner elliptic regions). Image reproduced and modified from Ref.~\cite{ESS-2}. Lower: (b) $\Delta\gamma^{112}$ and (c) $\Delta\gamma^{132}$ as a function of $v_2$ with events categorized by $q_{2,{\rm PPOI}}^2$ in the 30\%--40\% centrality range of Au+Au collisions at $\sqrt{s_{NN}} = 14.6$ GeV. POI are charged hadrons, excluding (anti)protons. The ESS technique extrapolates an observable to isotropic emission, characterized by zero elliptic flow ($v_2$). The error bars are statistical only. Linear fits (dashed lines) are used to extract the $y$-intercepts.
}
\label{fig2}
\end{figure}
\begin{figure*}[!tbhp]
\centering
\includegraphics[width=\textwidth]{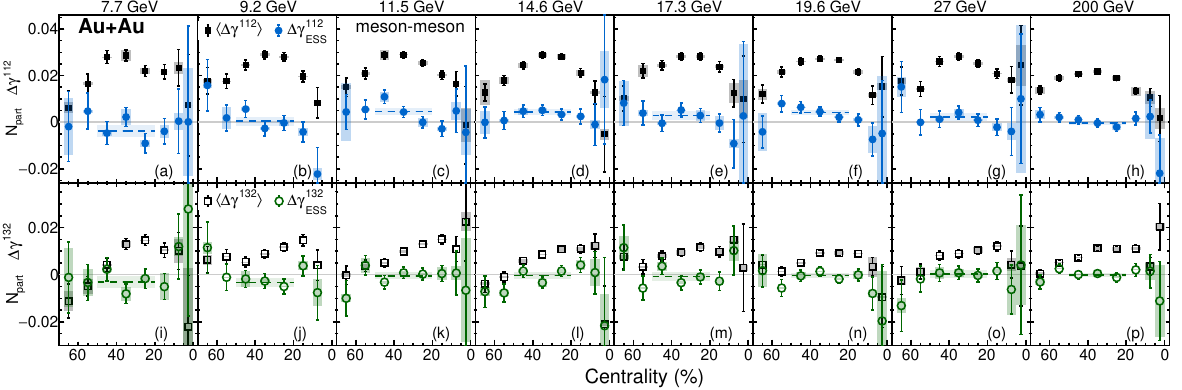}
\caption{Centrality dependence of (a--h) $N_{\rm part}\Delta\gamma^{112}$ and (i--p) $N_{\rm part}\Delta\gamma^{132}$ in Au+Au collisions at $\sqrt{s_{NN}}$= 7.7--200 GeV. Both the ensemble averages (squares) and the ESS results (circles) are presented. The error bars and boxes represent the statistical and systematic uncertainties, respectively. The dashed lines represent constant fits over the 20\%--50\% centrality range. The shaded bands denote the fit uncertainties. 
}
\label{fig3}
\end{figure*}

Figure~\ref{fig2}(a) illustrates that the emission pattern varies considerably for events with the same eccentricity~\cite{berndt-2}.
The ESS technique aims to project the charge separation observable towards isotropically emitting events, requiring the event shape variable to come from the same kinematic region as POI to better incorporate short-range fluctuations~\cite{ESS-2}.
To avoid the self-correlation between  $v_{2}$ and the event shape variable, we construct the latter using particle pairs of interest (PPOI) instead of individual particles~\cite{ESS-2}, 
\begin{equation}
q^2_{2,{\rm PPOI}}= \frac{\bigl(\sum^{N_{\rm pair}}_{i=1} \sin2\varphi_i^{\rm p}\bigr)^2 + \bigl(\sum^{N_{\rm pair}}_{i=1} \cos2\varphi_i^{\rm p}\bigr)^2 }{N_{\rm pair}(1+N_{\rm pair} v_{2, \rm pair}^2)},    
\end{equation}
where $N_{\rm pair}$ is the number of PPOI in one event. $\varphi_i^{\rm p}$ denotes the azimuthal angle of the momentum sum for a pair, and $v_{2, \rm pair} \equiv \langle \cos2(\varphi^{\rm p} - \Psi_{\rm{RP}})\rangle$ is averaged over all PPOI, regardless of charge, and over all events.
We first use $q_{2,\rm{PPOI}}^2$ to categorize events into different subsets $j$, and in each subset, measure $\Delta\gamma^{112}_j$ and $v_{2,j}$. 
Then, we plot $\Delta\gamma^{112}_j$ against $v_{2,j}$, as illustrated in Fig.~\ref{fig2}(a), and perform a linear fit to obtain the $y$-intercept at zero $v_2$.
Finally, we apply a small correction $(1- \overline{v}_2)^2$ to the $y$-intercept to recover the unbiased signal $\Delta\gamma^{112}_{\rm{ESS}}$~\cite{1st-ESS,ESS-1,ESS-2,non-inter-dept}, where $\overline{v}_2$ is the average over all subsets.

We analyzed 60 million events from Au+Au collisions at $\sqrt{s_{NN}} = 7.7$ GeV (2021), 110 million at 9.2 GeV (2019--2020), 160 million at 11.5 GeV (2020), 230 million at 14.6 GeV (2019), 200 million at 17.3 GeV (2021), 360 million at 19.6 GeV (2019), 490 million at 27 GeV (2018), and 2 billion at 200 GeV (2016).
The STAR time projection chamber (TPC)~\cite{TPC-1} reconstructs charged-particle helices and determines the primary vertex position of each event along the beam direction ($V_{z,{\rm TPC}}$) and its radial distance from the $z$ axis ($V_r$).
We selected events with $|V_{z,{\rm TPC}}| < 70$ cm (30 cm for 200 GeV) and $|V_r| < 2 $ cm. 
At 200 GeV, we also required $|V_{z,{\rm TPC}}-V_{z,{\rm VPD}}| <$ 3 cm, where $V_{z,{\rm VPD}}$ was measured by the Vertex Position Detector (VPD)~\cite{VPD}. 
Events chosen with minimum bias triggers are categorized into centrality classes according to the observed charged-particle multiplicity at midrapidities, following the procedure outlined in Ref.~\cite{STAR-11}.
Outlier events showing multiplicity correlation discrepancies between the TPC and the fast time-of-flight detector~\cite{tof} are excluded as out-of-time pile-up events.

The POI selection involves a cut on the number of ionization points used in the helix reconstruction ($N_{\rm hits} \ge 15$) to ensure track quality, and another cut on the distance of the closest approach to the primary vertex (DCA $< 3$ cm) to suppress contributions from weak decays and secondary interactions.
The kinematic cuts on pseudorapidity, momentum, and transverse momentum are  $|\eta| < 1$, $p < 1.4$ GeV/$c$, and $p_T > 0.2$ GeV/$c$, respectively.
Given the pronounced disparity in collective motions between protons and antiprotons at BES energies~\cite{STAR-BESv1,STAR-BESv2} and the different CME expectations for $p$ and $\pi/K$~\cite{CME-5}, we omit (anti)protons from POI by requiring $n\sigma_{\rm p}< -2$. 
Here, $n\sigma_{\rm p}$ quantifies the difference, in terms of standard deviation, between the measured ionization energy loss in the TPC and its expected value for protons~\cite{nSigma1,nSigma2}.
At $\sqrt{s_{NN}}$=7.7 GeV, where protons are more abundant, we tighten the cuts to $p < 1.3$ GeV/$c$ and $n\sigma_{\rm p}< -2.5$.

The STAR event plane detector (EPD, $2.1<|\eta|<5.1$)~\cite{EPD-1} extends into the forward/backward region that is rich in spectator protons at BES-II energies. 
We construct the spectator plane $\Psi^f$ ($\Psi^b$) at forward (backward) rapidities using the EPD hits at $|\eta|$ above the corresponding beam rapidity for collisions at $\sqrt{s_{NN}} =$ 7.7--27 GeV, and using the zero degree calorimeter shower-maximum detectors (ZDC-SMD)~\cite{ADLER2001488} for collisions at 200 GeV.
The detailed procedures follow those adopted in Refs.~\cite{STAR-27-CME-1, STAR-v1-62GeV}.
The spectator plane exhibits a stronger correlation with the magnetic field direction than the participant plane~\cite{TwoPlane1,TwoPlane2} and minimizes the nonflow backgrounds~\cite{STAR-8,STAR-27-CME-1}.
We measure three observables using the spectator planes, 
\begin{eqnarray}
v_2 &=& \langle \cos(2\varphi-\Psi^f-\Psi^b)\rangle/ \langle \cos(\Psi^f-\Psi^b)\rangle, \label{eq2}\\
\gamma^{112} &=& \langle \cos(\varphi_\alpha+\varphi_\beta-\Psi^f-\Psi^b)\rangle / \langle \cos(\Psi^f-\Psi^b)\rangle, \label{eq3}\\
\gamma^{132} &=& \langle \cos(\varphi_\alpha-3\varphi_\beta+\Psi^f+\Psi^b)\rangle / \langle \cos(\Psi^f-\Psi^b)\rangle.~~~\label{eq4}
\end{eqnarray}
The denominator in Eqs.~(\ref{eq2}), (\ref{eq3}), and (\ref{eq4}) compensates for the finite event plane resolution.
We employ $\gamma^{132}$~\cite{Subikash} as a background indicator because it is subject to similar flow-related background mechanisms as $\gamma^{112}$~\cite{ESS-1, accompany-long-paper}. 

The procedure to assess systematic uncertainties is similar to those used in Ref.~\cite{Isobar-1}. 
We estimate the systematic variation in the observables by making changes in each of the following cuts, $0 < V_z < 70$ cm ($0 < V_z < 30$ cm for 200 GeV), $N_{\rm hits}\ge20$, DCA $< 1$ cm, and $n\sigma_{\rm p}<-3$ ($-3.5$ for 7.7 GeV). 
Additionally, split tracks are rejected by requiring each accepted track to have the ratio of $N_{\rm hits}$ to the maximum possible number of ionization points greater than 0.52.  
For each of these cuts, we measure the difference $\Delta_i$ between the results with the default cuts and with the varied cuts.
For each variation $i$, we use the Barlow method~\cite{Barlow-1} to account for the impact of statistics.
Let $\sigma_{{\rm stat},d}$ and $\sigma_{{\rm stat},i}$ denote the statistical uncertainties for the results with the default and varied cuts, respectively.
If $\Delta_i^2$ exceeds $|\sigma^2_{{\rm stat},i} - \sigma^2_{{\rm stat},d}|$, we define $\sigma_i = \sqrt{\Delta_i^2 - |\sigma^2_{{\rm stat},i} - \sigma^2_{{\rm stat},d}|}$.
Otherwise, $\sigma_i = 0$. 
The overall systematic uncertainty is the quadrature sum of $\sigma_i/\sqrt{12}$ (or $\sigma_i/\sqrt{3}$ for the $V_z$ variation), assuming the default and varied-cut results constrain the maximum and minimum, regardless of order, under a flat prior~\cite{FlatPrior} for each source. The final central value is taken as the mean of the averages of these extremes across all five sources~\cite{accompany-long-paper}.
The DCA and $n\sigma_{\rm p}$ cuts are the primary sources of systematic error, each typically at the level of 10\% of the corresponding statistical uncertainty.
The variations in the rest of the cuts have negligible effects.

Figures~\ref{fig2}(b) and~\ref{fig2}(c) demonstrate the ESS approach, depicting $\Delta\gamma^{112}$ and $\Delta\gamma^{132}$, respectively, vs $v_2$ with events classified by $q_{2,{\rm PPOI}}^2$ in the 30\%--40\% centrality range of Au+Au collisions at $\sqrt{s_{NN}} = 14.6$ GeV.
The lowest $v_2$ value is close to zero, facilitating the projection to the zero-flow limit. 
We restore the unbiased signals with $\Delta\gamma^{112(132)}_{\rm ESS} = (1- \overline{v}_2)^2\Delta\gamma^{112(132)}|_{v_{2} = 0}.$

Figure~\ref{fig3} shows the centrality dependence of (a--h) $N_{\rm part}\Delta\gamma^{112}$ and (i--p) $N_{\rm part}\Delta\gamma^{132}$ in Au+Au collisions at $\sqrt{s_{NN}}$= 7.7--200 GeV. 
$N_{\rm part}$ is the number of participating nucleons~\cite{accompany-long-paper}.
For both observables, the ESS results (circles) exhibit a substantial reduction compared with the ensemble averages (squares).
While $N_{\rm part}\Delta\gamma^{132}_{\rm ESS}$ aligns with zero in all centrality intervals at all beam energies, $N_{\rm part}\Delta\gamma^{112}_{\rm ESS}$ is generally finite in mid-central collisions between 10 and 20 GeV.
We perform constant fits over the 20\%--50\% centrality range to reduce statistical uncertainties.

\begin{figure}[tb]
\centering
\includegraphics[width=0.48 \textwidth]{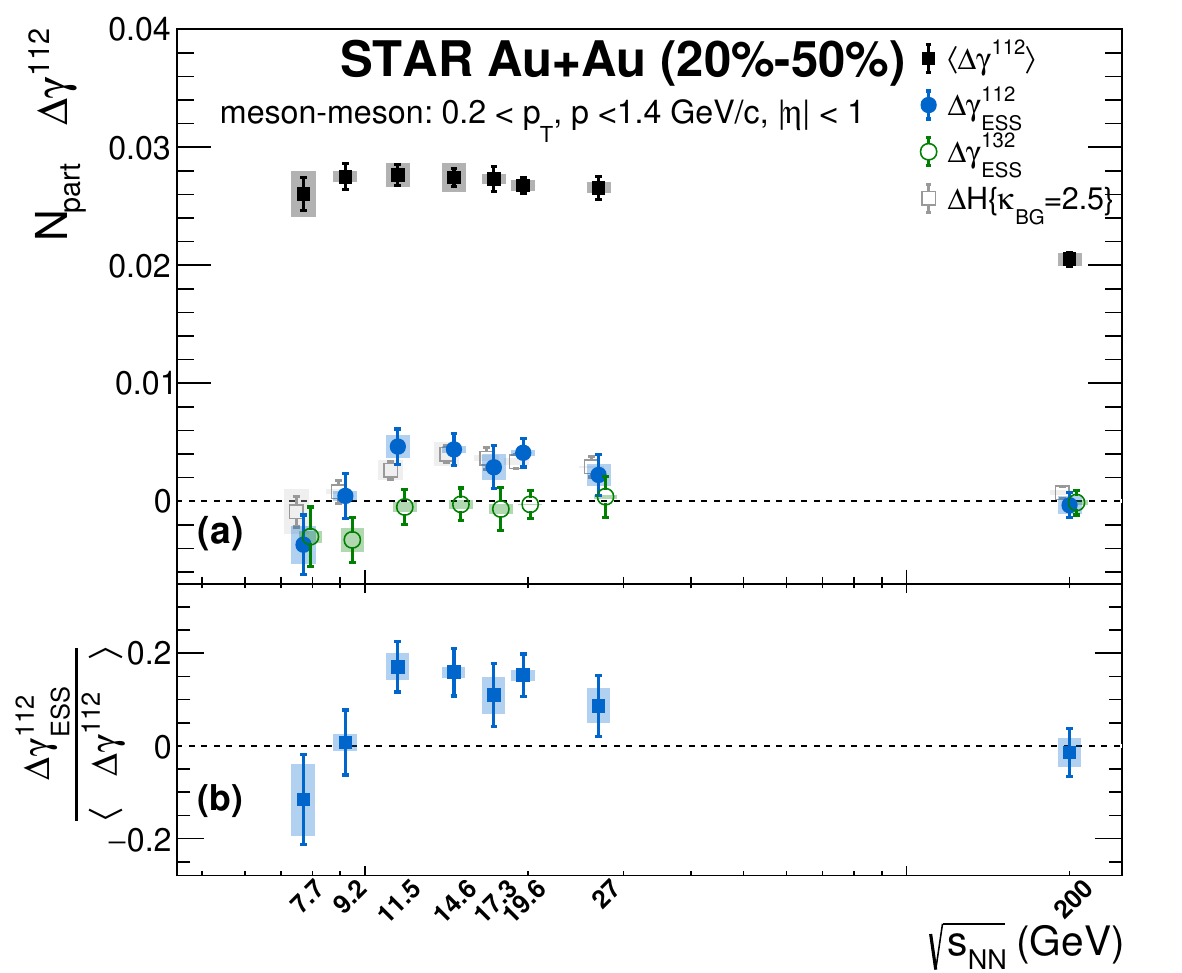}
\caption{(a) Beam-energy dependence of $N_{\rm part} \langle\Delta\gamma^{112}\rangle$ (full square), $N_{\rm part}\Delta\gamma^{112}_{\rm{ESS}}$ (full circles), and $N_{\rm part}\Delta\gamma^{132}_{\rm{ESS}}$ (open circles) integrated over the 20\%--50\% centrality range in Au+Au collisions. For comparison, we also add the results for $N_{\rm part}\Delta H(\kappa_{bg}=2.5)$ (open square). 
(b) The corresponding ratio of $\Delta\gamma^{112}_{\rm{ESS}}$ to $\langle\Delta\gamma^{112}\rangle$. Some points are slightly shifted horizontally to enhance clarity. The error bars and boxes represent the statistical and systematic uncertainties, respectively.}
\label{fig4}
\end{figure}

Figure~\ref{fig4}(a) presents the beam-energy dependence of $N_{\rm part} \langle\Delta\gamma^{112}\rangle$, $N_{\rm part}\Delta\gamma^{112}_{\rm{ESS}}$, and $N_{\rm part}\Delta\gamma^{132}_{\rm{ESS}}$ integrated over the 20\%--50\% centrality range in Au+Au collisions.
The background indicator $\Delta\gamma^{132}_{\rm{ESS}}$ is consistent with zero at all energies, affirming the effectiveness of flow-related background suppression with the ESS and nonflow background removal with the spectator plane. 
At $\sqrt{s_{NN}}$= 200 GeV,  $\Delta\gamma^{112}_{\rm{ESS}}$ is consistent with zero.
Similar null results have been reported in previous research in Au+Au collisions at this energy~\cite{STAR-8,STAR-6}. 
This helps explain why the isobar collisions at $\sqrt{s_{NN}}$= 200 GeV yield a null result.
At BES-II energies, the ensemble average $N_{\rm part} \langle\Delta\gamma^{112}\rangle$ remains relatively constant. 
After the background subtraction, $N_{\rm part}\Delta\gamma^{112}_{\rm{ESS}}$ reveals a discernible and statistically significant charge separation signal
at $\sqrt{s_{NN}}$= 11.5, 14.6, and 19.6 GeV, with significance levels of 2.6$\sigma$, 3.1$\sigma$, and 3.3$\sigma$, respectively.
The remaining charge separation signals are consistent with zero at $\sqrt{s_{NN}}$= 7.7 and 9.2 GeV.
The significance levels at $\sqrt{s_{NN}}$= 17.3 and 27 GeV are $1.3\sigma$ and $1.1\sigma$, repsectively.

Assuming comparable physics conditions between $\sqrt{s_{NN}}$= 10 and 20 GeV, the statistically-weighted average of $N_{\rm part}\Delta\gamma^{112}_{\rm ESS}$  over $\sqrt{s_{NN}}$= 11.5, 14.6, 17.3, and 19.6 GeV is $[4.11 \pm 0.71(stat.) \pm 0.23(syst.)] \times 10^{-3}$, yielding a $5.5\sigma$ significance of the charge separation signal. 
In addition, we find a good agreement between $N_{\rm part}\Delta\gamma^{112}_{\rm{ESS}}$ and $N_{\rm part}\Delta H(\kappa_{bg} = 2.5)$, indicating that the background mechanism can be described with a universal coupling between $v_2$ and $\Delta\delta\equiv\delta_{\rm OS} - \delta_{\rm SS}$.

Figure~\ref{fig4}(b) shows the ratio of $\Delta\gamma^{112}_{\rm{ESS}}$ to $\langle\Delta\gamma^{112}\rangle$, calculated for each centrality interval from Fig.~\ref{fig3} and then averaged over the 20\%--50\% centrality range.
The ESS method identifies that at least 80\% of $\langle\Delta\gamma^{112}\rangle$ arises from the flow-related background.
The ratio $\Delta\gamma^{112}_{\rm{ESS}}/\langle\Delta\gamma^{112}\rangle$ exhibits similar significance levels ($2.8\sigma$, $3\sigma$, and $3.2\sigma$) as $N_{\rm part}\Delta\gamma^{112}_{\rm ESS}$ at $\sqrt{s_{NN}}$= 11.5, 14.6, and 19.6 GeV.
The ratio at $\sqrt{s_{NN}}$= 200 GeV is $-0.02\pm0.05(stat.)\pm 0.03(syst.)$, compatible with the upper limit of the CME fraction in $\Delta\gamma^{112}$ ($f_{\rm CME}$) reported for isobar collisions~\cite{Isobar-2} and Au+Au collisions~\cite{STAR-8,STAR-6} at $\sqrt{s_{NN}}$= 200 GeV.
At $\sqrt{s_{NN}}$= 27 GeV, the ratio is consistent with the upper limit of $f_{\rm CME}$ estimated from a prior STAR measurement~\cite{STAR-27-CME-1}.

The persistence of the charge separation signal $\Delta\gamma^{112}_{\rm ESS}$ in mid-central Au+Au collisions at beam energies between $\sqrt{s_{NN}}$= 10 and 20 GeV, and its absence elsewhere at lower and higher beam energies, warrant additional theoretical studies for these energies.
For example, QCD topological vacuum transitions could be amplified near the critical-point region~\cite{Kharzeev-critical}.
Additionally, the time-dependent dynamics of the magnetic field may favor low $\sqrt{s_{NN}}$~\cite{LHui-Bfield,HuangAnping-Bfield}, where the intense magnetic field could persist 
long enough to penetrate the QGP phase. At higher beam energies, the magnetic field may drop to a significantly weaker strength in the QGP phase~\cite{DynamicalEvolution}.
Below $\sqrt{s_{NN}}$= 10 GeV, chiral symmetry restoration may be significantly diminished, even in the presence of possible partonic degrees of freedom~\cite{Mizher:2010zb}, causing the precondition for the CME to vanish.

In summary, we have presented measurements of charge separation correlations along the magnetic field direction using Au+Au collisions at RHIC from $\sqrt{s_{NN}}$= 7.7 to 200 GeV energies, with the flow-related background effectively suppressed.
We report a remaining charge separation signal 
in mid-central Au+Au collisions, positive finite with around $3\sigma$ significance at each of the center-of-mass energies of $\sqrt{s_{NN}}$= 11.5, 14.6, and 19.6 GeV. 
The results at $\sqrt{s_{NN}}$= 17.3 and 27 GeV also show positive values but with a lower significance of $1.3\sigma$ and $1.1\sigma$. 
Below $\sqrt{s_{NN}}$= 10 GeV or at $\sqrt{s_{NN}}$= 200 GeV, the charge separation is consistent with zero. 
When the data between $\sqrt{s_{NN}}$= 10 and 20 GeV are combined, the significance rises to $5.5\sigma$. 
The absence of a definitive CME signal from the top RHIC energy and the LHC energies~\cite{CMS-2, ALICE-fcme} can constrain the dynamical evolution of the magnetic field in the QGP phase in these collisions.
Our measurements call for more investigation into the magnetic field evolution in a QGP and the QCD topological vacuum transitions at lower RHIC energies.

\begin{acknowledgments}{
We thank the RHIC Operations Group and SDCC at BNL, the NERSC Center at LBNL, and the Open Science Grid consortium for providing resources and support.  This work was supported in part by the Office of Nuclear Physics within the U.S. DOE Office of Science, the U.S. National Science Foundation, National Natural Science Foundation of China, Chinese Academy of Science, the Ministry of Science and Technology of China and the Chinese Ministry of Education, NSTC Taipei, the National Research Foundation of Korea, Czech Science Foundation and Ministry of Education, Youth and Sports of the Czech Republic, Hungarian National Research, Development and Innovation Office, New National Excellency Programme of the Hungarian Ministry of Human Capacities, Department of Atomic Energy and Department of Science and Technology of the Government of India, the National Science Centre and WUT ID-UB of Poland, the Ministry of Science, Education and Sports of the Republic of Croatia, German Bundesministerium f\"ur Bildung, Wissenschaft, Forschung and Technologie (BMBF), Helmholtz Association, Ministry of Education, Culture, Sports, Science, and Technology (MEXT), Japan Society for the Promotion of Science (JSPS) and Agencia Nacional de Investigaci\'on y Desarrollo (ANID) of Chile.
}
\end{acknowledgments}

{}
\end{document}